\documentstyle[aps,psfig,floats]{revtex}

\begin{document}
\draft
\title{Stochastic resonance in a suspension of magnetic dipoles under shear  flow.}
\author{Tom\'as Alarc\'on and Agust\'{\i}n P\'erez-Madrid}
\address{Departament de F\'{\i}sica Fonamental and CER on Physics of Complex Systems\\
Facultat de F\'{\i}sica\\
Universitat de Barcelona\\
Diagonal 647, 08028 Barcelona, Spain\\
}
\maketitle

\begin{abstract}
We show that a magnetic dipole in a shear  flow under the action of an oscillating magnetic field displays stochastic resonance in the linear response regime. To this end, we compute the classical quantifiers of stochastic resonance, \emph{i.e.} the signal to noise ratio, the escape time distribution, and the mean first passage time. We also discuss limitations and role of the linear response theory in its applications to the theory of stochastic resonance.
\end{abstract}

\pacs{Pacs: 05.40.+a}

\section{introduction}

The dynamics of periodically-driven stochastic systems has been an active field of research over the last years \cite{kn:jung}. This kind of systems arise commonly in fields of physics, chemistry and biology. Examples are found in problems involving transport at the cellular level \cite{kn:magn,kn:astumian}, optical and electronic devices \cite{kn:mcwr} and signal transduction in neuronal tissue \cite{kn:longtin,kn:vsr}, to cite just a few.

A particularly interesting phenomenon, occurring in periodically-driven non-linear noisy systems, is stochastic resonance (SR) \cite{kn:ghjm}. This term refers to the enhancement of the response of the system to a coherent signal when the intensity of the noise grows, instead of the degradation that one naively expects. The mechanism leading to this phenomenon is quite simple. Imagine a system which exhibits an energetic activation barrier. In the presence of noise, the system can be assumed to surmount this barrier with a rate proportional to $e^{-\Delta E/D}$, where $\Delta E$ is the height of the barrier and D is the intensity of the noise acting on the system. The inverse of this rate defines the average waiting time, $T(D)$, between two noise-induced transitions. In the presence of a periodic forcing, the height of the barrier is periodically raised an lowered. When the period of the external force synchronizes with $2T(D)$, the barrier surmounting will be enhanced by the cooperative effect of the noise and the periodic forcing.

Although originally proposed for systems in a double-well potential, this original scheme has been overcome. In fact, it is known that SR is exhibited by several classes of monostable systems, among which one could mention excitable and threshold systems \cite{kn:wiesenfeld,kn:kiss,kn:cbg,kn:vgr} or systems which do not follow an activated dynamics but a relaxational dynamics \cite{kn:dykman,kn:vilar}. 

In this paper we will show that a magnetic dipole immerse in a shear flow exhibits stochastic resonance when a weak oscillating magnetic field is acting on it. The presence of this flow takes the system out of equilibrium impressing certain peculiarities in the behaviour of this system. In order to treat this problem we will analyze the response of the system in the linear regime. Previous study of the dynamics of a dipole under an oscillating magnetic field has revealed that linear response theory (LRT) predicts a monotonically decreasing behaviour for the ratio between the output signal and the output noise or signal to noise ratio (SNR), \emph{i.e.} for very weak applied fields the dipole does not exhibit SR \cite{kn:vilar}. In the present case there is a new ingredient, absent in \cite{kn:vilar}: the presence of the shear flow, which is determinant for many interesting aspects of the dynamics of this system. Additionally, although we show the existence of SR in the linear regime, we discuss limitations and role of the LRT in its application to the theory of SR, mainly related to questions concerning the fluctuation-dissipation theorem.

The paper is organized as follows: in section II we analyze the dynamics of a  dipole in a shear flow and find the fixed points. Section III is devoted to study the response of the system to an oscillating magnetic field by computing the susceptibility. In section IV we calculate the power spectrum and the signal-to-noise ratio. In section V we compute the escape time distribution and from it the mean first passage time. Finally, in section VI we discuss our results.

\section{Dynamics of a dipole in a shear flow: fixed points and their stability}

We consider a dilute colloidal suspension of ferromagnetic dipolar spherical
particles, with magnetic moment, $\vec{m}= m_{s}\hat{
\vec{R}}$, where $\hat{\vec{R}}$ is an unit vector accounting for the
orientation of the dipole, therefore, being the magnetic moment rigidly attached to the particles. Each dipole is under the influence of a shear flow with vorticity $\vec{\Omega}= 2\omega _{0}\hat{\vec{z}}$, with 
$\hat{\vec{z}}$ being the unit vector along the $z$-axis, and of an oscillating field $\vec{H}= He^{-i\omega t}\hat{\vec{x}}$, with $\hat{\vec{x}}$ being the 
unit vector along the $x$-axis. The dynamics of these dipoles is governed by the following equation of motion

\begin{equation}\label{2.39}
I\frac{d\vec{\Omega}_p}{dt}=\vec{m}\times\vec{H} + \xi_r\left(\frac{1}{2}\vec{\Omega} - \vec{\Omega
}_p\right),
\end{equation}

\noindent where $I$ is the moment of inertia of the particles, $\xi _{r}= 8\pi \eta_0 a^{3}$ is the rotational friction coefficient, with $\eta_0 $ the solvent viscosity, and $a$ the radius of the particle. For $t\gg \tau _{r}$,with $\tau _{r}= I/\xi _{r}$ being the inertial time scale, the motion of the particle enters in the overdamped regime.  This time scale defines a cut-off frequency $\omega _{r}= \tau _{r}^{-1}$, such that the condition for overdamped motion is equivalent to $\omega \ll \omega _{r}$. In this case eq. (\ref{2.39}) yields the balance condition between the magnetic and hydrodynamic torques acting on each particle

\begin{equation}  \label{2.40}
\vec{m}\times\vec{H} + \xi_r\left(\frac{1}{2}\vec{\Omega} - \vec{\Omega
}_p\right) = 0 ,
\end{equation}

\noindent which, together with the rigid rotor evolution equation

\begin{equation}  \label{2.41}
\frac{d \hat{\vec{R}}}{d t} = \vec{\Omega}_p\times\hat{\vec{R}},
\end{equation}

\noindent leads to the dynamic equation for $\hat{\vec{R}}$ 

\begin{equation}  \label{2.42}
\frac{d \hat{\vec{R}}}{d t} = \omega_0\left\{\hat{\vec{z}} \right.+ 
\lambda
(t)\left.\left (\hat{\vec{R}}\times\hat{\vec{x}}\right )\right\}\times\hat
{\vec{R}}.
\end{equation}

\noindent Here $\lambda (t)\equiv (m_{s}H/\xi _{r}\omega_{0})e^{-i\omega t}$, with $\vec{\Omega}_{p}$ being the angular velocity of the particle.

The computation of the fixed points of eq. (\ref{2.42}) when the magnetic field is held constant, \emph{i.e.} $\lambda(t)=\lambda_{0}=cnt.$, and their linear stability analysis are given in detail in Appendix A. After some algebra eq. (\ref{2.42}) becomes

\begin{equation}\label{2.43}
\frac{d \hat{\vec{R}}}{d t}=\omega_{0}\left[\hat{\vec{z}}\times\hat{\vec{R}}+\lambda\hat{\vec{x}}-\lambda\hat{\vec{R}}(\hat{\vec{R}}\cdot\hat{\vec{x}})\right].
\end{equation}

For $\lambda_{0}\geq 1$, this equation has only a linearly stable stationary state. The orientation of the suspended particles is fixed to 

\begin{equation}\label{2.44}
\hat{\vec{R}}_{s}=\sqrt{1-\lambda^{-2}}\hat{\vec{x}}+\lambda^{-1}\hat{\vec{y}}.
\end{equation}

\noindent This means that in this regime the hydrodynamic torque, which tends to cause the rotation of the particles is insufficient to overcome the magnetic torque which maintains constant their orientations.   

For $\lambda_{0}< 1$, which is the case we are interested in, the particles undergo a rotation around a fixed axis lying in the $y$-$z$ plane, being the director of this axis given by

\begin{equation}\label{2.45}
\hat{\vec{R}}_{s}=\pm\sqrt{1-\lambda^{2}}\hat{\vec{y}}+\lambda\hat{\vec{z}}.
\end{equation}

\noindent In this case the hydrodynamic torque is strong enough to make the dipole to precess around the orientation $\hat{\vec{R}}_{s}$ eq. (\ref{2.45}) (see Appendix A). 

\section{Response to an oscillating magnetic field}

The analysis of Section II has been carried out for the deterministic dynamics of a magnetic dipole in a shear flow. Fluctuations are introduced by means of a Brownian torque. The corresponding Langevin equation is

\begin{equation}\label{2.46}
\frac{d \hat{\vec{R}}}{d t} = \omega_0\left\{ \lambda(t)\left(\hat{\vec{R}}\times\hat{\vec{x}}\right)+\hat{\vec{z}}+\frac{1}{\xi_{r}\omega_{0}}\left(\hat{\vec{R}}\times\vec{F}_{B}(t)\right)\right\}\times\hat
{\vec{R}},
\end{equation}

\noindent where $\vec{F}_{B}(t)$ is a Gaussian white noise of zero mean and correlation function

\begin{equation}\label{2.47}
\langle\vec{F}_{B}(t)\vec{F}_{B}(t')\rangle=2\xi_{r}K_{B}T\delta(t-t').
\end{equation}

The Fokker-Planck equation corresponding to eq. (\ref{2.46}) is given by 

\begin{equation}  \label{2.48}
\partial_{t}\Psi(\hat{\vec{R}},t)\,=\,({\cal L}_{0}+\lambda(t){\cal L}
_{1})\Psi(\hat{\vec{R}},t),
\end{equation}

\noindent where ${\cal L}_{0}$ and ${\cal L}_{1}$ are operators defined by

\begin{eqnarray}  \label{2.49}
{\cal L}_{0}&=&-\omega_{0}\,\hat{\vec{z}}\cdot{\cal \vec{R}}+D_{r}{\cal
 \vec{%
R}}^{2},  \nonumber \\
{\cal L}_{1}&=&2\omega_{0}\hat{\vec{R}}\cdot\hat{\vec{x}}- 
\omega_{0}(\hat{\vec{R}}\times\hat{\vec{x}})\cdot{\cal \vec{R}},  
\end{eqnarray}

\noindent with $D_{r}=k_{B}T/\xi _{r}$ being the rotational diffusion
coefficient, and ${\cal \vec{R}}=\hat{\vec{R}}\times \partial /\partial
 \hat{%
\vec{R}}\;$ the rotational operator. Notice that the first and second terms
 on
the right hand side of eq. (\ref{2.49})$_{1}$, correspond to convective
and diffusive term, respectively. Moreover eq. (\ref{2.48}) which,  according to eq. (\ref{2.46}), rules the Brownian dynamics in the case of
overdamped motion, is valid in the diffusion regime. This regime is also
characterized by the condition $t\gg \tau _{r}$, or equivalently $\omega 
\ll
\omega _{r}$, which implicitly involves the white noise assumption.

To solve the Fokker-Planck equation (\ref{2.48}) we will assume that $%
\lambda _{0}\equiv \vert \lambda(t)\vert$ constitutes a small parameter such that this equation
can be solved perturbatively. Thus up to first order in $\lambda $, the
solution of the Fokker-Planck equation (\ref{2.48}) is

\begin{equation}  \label{2.50}
\Psi(\hat{\vec{R}},t)\,=\,e^{(t-t_{0}){\cal L}_{0}}\Psi^{0}(t_{0})+%
\int_{t_{0}}^{t}\,dt^{\prime}\,\lambda(t^{\prime})e^{(t-t^{\prime}){\cal 
L}%
_{0}}{\cal L}_{1}\Psi_{0}(t^{\prime})
\end{equation}

\noindent Here $\Psi_{0}(t^{\prime}) = e^{(t^{\prime}-t_0){\cal L}%
_{0}}\Psi^{0}(t=t_{0})$ is the zero order solution at time $t^{\prime}$
, and

\begin{equation}  \label{2.51}
\Psi^{0}(\hat{\vec{R}},t =t_0)\,=\,\delta(\hat{\vec{R}}-\hat{\vec{R}}_0) ,
\end{equation}

\noindent with $\hat{\vec{R}}_0$ being an arbitrary initial orientation. 
As follows from eq. (\ref{2.49})$_1$, the unperturbed operator ${\cal L
}
_{0}$ is composed of the operators ${\cal R}_{z}$ and ${\cal R}^{2}$, which
are proportional to the orbital angular momentum operators of quantum
mechanics $L_{z}$ and $L^{2}$, respectively, and, therefore, their
eigenfunctions are the spherical harmonics\cite{kn:sak}

\begin{eqnarray}  \label{2.52}
{\cal R}_{z}Y_{l\, m}(\hat{\vec{R}})&=& imY_{l\, m}(\hat{\vec{R}}), \nonumber \\
{\cal R}^{2}Y_{l\, m}(\hat{\vec{R}})&=& -l(l+1)Y_{l\, m}(\hat{\vec{R}}) .
\end{eqnarray}

Given that we know how $\vec{{\cal R}}$ acts on the spherical harmonics, 
it
is convenient to expand the initial condition in series of these functions,
since the spherical harmonics constitute a complete set of functions which
are a basis in the Hilbert space of the integrable functions over the unit
sphere\cite{kn:ch}

\begin{equation}  \label{2.53}
\Psi^{0}(\hat{\vec{R}},t_0)\, =\,\delta(\hat{\vec{R}}-\hat{\vec{R}}_0)\,
=\,\sum_{l=0}^{\infty}\,\sum_{m=-l}^{l}\, Y_{l\, m}^{*}(\hat{\vec{R
}}%
_0)Y_{l\, m}(\hat{\vec{R}} )\;\; .
\end{equation}

\noindent Using this expansion in eq. (\ref{2.50}), for the first order
correction to the probability density, $\Delta\Psi\equiv\Psi-\Psi_{0}$, we obtain

\begin{equation}  \label{2.54}
\Delta\Psi(\hat{\vec{R}},t)\,
=\,\sum_{l=0}^{\infty}\;\sum_{m=-l}^{l}\;\int_{t_{0}}^{t}\,
dt^{\prime}\lambda(t^{\prime})Y_{l\, m}^{*}(\hat{\vec{R}}_0)e^{(t-t^{\prime})%
{\cal L}_0}{\cal L}_1 e^{(t^{\prime}-t_0){\cal L}_{0}}Y_{l\, m}(\hat{\vec
{R}}%
) .
\end{equation}

\noindent Notice that the integral of $\Delta\Psi(\hat{\vec{R}},t)$ over 
the
entire solid angle is zero, in agreement with the fact that the unperturbed
solution $\Psi_0(\hat{\vec{R}},t)$ is normalized.

Since, we are interested in the asymptotic behavior we will set $
t_{0}\rightarrow -\infty$. In this limit, eq. (\ref{2.50}) becomes

\begin{equation}  \label{2.55}
\Psi (\hat{\vec{R}} ,t) = \frac{1}{4\pi}\left\{ 1 + \int_{-\infty}^{t}\right.\;
dt^{\prime}\,\lambda (t^{\prime })\,e^{(t-t^{\prime }){\cal L}_{0}}2
\left.\right.
\left. \hat{\vec{R}}\cdot\hat{\vec{x}}\right\},
\end{equation}

\noindent where now

\begin{equation}  \label{2.56}
\Delta \Psi (\hat{\vec{R}},t)=\frac{1}{4\pi }\int_{-\infty}^{t} \;dt^{\prime}\,\lambda (t^{\prime })\,e^{(t-t^{\prime }){\cal L}_{0}}\,2
\hat{
\vec{R}}\cdot \hat{\vec{x}},
\end{equation}
\noindent and 

\begin{equation}  \label{2.57}
\Psi ^{0}(\hat{\vec{R}},t)=\frac{1}{4\pi}
\end{equation}

\noindent is the uniform distribution function in the unit sphere.

From eq. (\ref{2.56}) the contribution of the AC field to the mean value of
the orientation vector $\hat{\vec{R}}$ can be obtained as

\begin{equation}  \label{2.58}
\hat{\vec{R}}(t)\,=\,\int\, d\hat{\vec{R}}\, \hat{\vec{R}} \,\Delta\Psi
\,
=\, \frac{1}{4\pi}\int_{-\infty}^{t}\;dt^{\prime}\lambda(t^{ \prime})\int\, d%
\hat{\vec{R}}\, \hat{\vec{R}}\, e^{(t-t^{\prime}){\cal L} _0}\, 2
\hat{\vec{R}}\cdot\hat{\vec{x}}.
\end{equation}

\noindent This equation can be written in the more compact form

\begin{equation}\label{2.59}
\hat{R}_{i}(t)\,=\,\int_{-\infty }^{t}\,dt^{\prime }\lambda (t^{\prime
})\chi _{i}(t-t^{\prime }), 
\end{equation}

\noindent where the response function\cite{kn:rdl} has been
defined as

\begin{equation}  \label{2.60}
\chi _{i}(\tau )\,=\,\frac{1}{4\pi }\int \,d\hat{\vec{R}}\,\hat{R}%
_{i}e^{\tau {\cal L}_{0}}\,2\hat{\vec{R}}\cdot \hat{\vec{x}},
\end{equation}

\noindent for $\tau >0$.

By causality, we can write $t\rightarrow \infty $ in the upper limit of the
integral in eq. (\ref{2.59}); hence, this equation becomes

\begin{equation}\label{2.61}
\hat{R}_{i}(t)\,=\,\chi _{i}(\omega )\lambda (t).  
\end{equation}

\noindent where $\chi _{i}(\omega )$ is the generalized susceptibility,
 which is the Fourier transform of $\chi _{i}(\tau )\,$ 

\begin{equation}  \label{2.62}
\chi _{i}(\omega )\,=\,\frac{1}{4\pi }\int_{-\infty }^{\infty }d\tau
e^{i\omega \tau }\int d\hat{\vec{R}}\hat{R}_{i}e^{\tau {\cal L}_{0}}2 \hat{\vec{R}}\cdot \hat{\vec{x}}
\end{equation}

\noindent From this equation we obtain the components of the susceptibility

\begin{eqnarray}  \label{2.63}
\chi _{x}(\omega )\,=\, &\frac{1}{3}\left\{ \left[ \frac{2D_{r}}{%
4D_{r}^{2}+(\omega -\omega _{0})^{2}}-i\frac{(\omega_{0}-\omega )}{%
4D_{r}^{2}+(\omega -\omega _{0})^{2}}\right] +\right. &  \nonumber \\ \vspace{.25cm}
&\left. \left[ \frac{2D_{r}}{4D_{r}^{2}+(\omega +\omega _{0})^{2}}+i\frac
{%
(\omega _{0}+\omega )}{4D_{r}^{2}+(\omega +\omega _{0})^{2}}\right] \right\}
&
\end{eqnarray}

\begin{eqnarray}  \label{2.64}
\chi_y(\omega)\, =\,&\frac{1}{3}\left\{\left[\frac{(\omega_0
-\omega)}{4D_r^2 + (\omega-\omega_0)^2}+i\frac{2D_r}{4D_r^2 +
(\omega-\omega_0)^2}\right]+\right .&  \nonumber \\ \vspace{.25cm}
&\left .\left[\frac{(\omega_0 +\omega)}{4D_r^2 + (\omega+\omega_0)^2}-i\frac{%
2D_r}{4D_r^2 + (\omega+\omega_0)^2}\right]\right\}&
\end{eqnarray}

\begin{equation}  \label{2.65}
\chi_z(\omega)\, =\, 0
\end{equation}

\noindent The quantities $\chi_x$, and $\chi_y$, have poles at $\omega = \pm\omega_0 \pm 2D_ri$. The inverse of the imaginary part of these poles, $(2D_r)^{-1}$, defines the Brownian relaxation time.

\section{Power spectrum}

In order to discern whether or not SR is present in the relaxation process of the system under consideration we compute the power spectrum, which, following the Wiener-Khinchine theorem, is given  by the Fourier transform of the correlation function \cite{kn:vkam2,kn:jung}. Since we will take as output signal the projection of $\hat{\vec{R}}$ parallel to the magnetic field, \emph{i.e.} $\hat{R}_{x}$, we only compute the correlation function of this quantity. 

The correlation function of $\hat{R}_{x}$ is defined by

\begin{equation}\label{2.66}
\langle\hat{R}_{x}(t)\hat{R}_{x}(t+\tau)\vert\hat{\vec{R}}_{0}(t_{0})\rangle=\int\,d\hat{\vec{v}}\int\,d\hat{\vec{u}}\,\hat{u}_{x}\hat{v}_{x}\Psi(\hat{\vec{v}},t;\hat{\vec{u}},t+\tau\vert\hat{\vec{R}}_{0},t_{0}),
\end{equation}

\noindent where the initial condition is taken as $\Psi(\hat{\vec{R}},t_{0})=\delta(\hat{\vec{R}}-\hat{\vec{R}}_{0})$. The above quantity can be calculated from the solution Fokker-Planck equation simply by recalling the following properties of a Markov process \cite{kn:vkam2}.

\begin{eqnarray}\label{2.67}
\nonumber\Psi(\hat{\vec{v}}_{1},t_{1};...;\hat{\vec{v}}_{n},t_{n})&=&\Psi(\hat{\vec{v}}_{1},t_{1})\Psi(\hat{\vec{v}}_{1},t_{1}\vert\hat{\vec{v}}_{2},t_{2};...;\hat{\vec{v}}_{n},t_{n}),\\
\Psi(\hat{\vec{v}}_{1},t_{1}\vert\hat{\vec{v}}_{2},t_{2};...;\hat{\vec{v}}_{n},t_{n})&=&\Psi(\hat{\vec{v}}_{1},t_{1}\vert\hat{\vec{v}}_{2},t_{2}).
\end{eqnarray}

\noindent where $t_{1}>t_{2}>...>t_{n}$. By combination of these two properties eq. (\ref{2.66}) becomes

\begin{equation}\label{2.68}
\langle\hat{R}_{x}(t)\hat{R}_{x}(t+\tau)\vert\hat{\vec{R}}_{0}(t_{0})\rangle=\int\,d\hat{\vec{v}}\hat{v}_{x}\Psi(\hat{\vec{v}},t\vert\hat{\vec{R}}_{0},t_{0})\int\,d\hat{\vec{u}}\,\hat{u}_{x}\Psi(\hat{\vec{u}},t+\tau\vert\hat{\vec{v}},t).
\end{equation}

To proceed further, we compute in first place the integral over $\hat{\vec{u}}$ in eq. (\ref{2.68}). From eq. (\ref{2.50}), if $\tau>0$ 

\begin{eqnarray}\label{2.69}
\nonumber\Psi(\hat{\vec{u}},t+\tau\vert\hat{\vec{v}},t)&=&e^{\tau{\cal L}_{0}}\delta(\hat{\vec{u}}-\hat{\vec{v}})+\int_{t}^{t+\tau}ds\,\lambda(s)e^{(t+\tau-s){\cal L}_{0}}{\cal L}_{1}e^{s{\cal L}_{0}}\delta(\hat{\vec{u}}-\hat{\vec{v}})\\
&=&\Psi_{0}(\hat{\vec{u}},t+\tau\vert\hat{\vec{v}},t)+\Delta\Psi(\hat{\vec{u}},t+\tau\vert\hat{\vec{v}},t),
\end{eqnarray}

\noindent thus we have

\begin{equation}\label{2.70}
\int\,d\hat{\vec{u}}\,\hat{u}_{x}\Psi(\hat{\vec{u}},t+\tau\vert\hat{\vec{v}},t)=
\int\,d\hat{\vec{u}}\,\hat{u}_{x}\Psi_{0}(\hat{\vec{u}},t+\tau\vert\hat{\vec{v}},t)+\int\,d\hat{\vec{u}}\,\hat{u}_{x}\Delta\Psi(\hat{\vec{u}},t+\tau\vert\hat{\vec{v}},t).
\end{equation}

\noindent The result of these integrals are (for the detailed derivation, see Appendix B)

\begin{eqnarray}\label{2.71}
\nonumber&&\int\,d\hat{\vec{u}}\,\hat{u}_{x}\Psi_{0}(\hat{\vec{u}},t+\tau\vert\hat{\vec{v}},t)=-\sqrt{\frac{2\pi}{3}}\left[e^{-(2D_{r}-i\omega_{0})\tau}Y_{11}(\hat{\vec{v}})+e^{-(2D_{r}+i\omega_{0})\tau}Y_{1-1}(\hat{\vec{v}})\right],\\
\nonumber&&\int\,d\hat{\vec{u}}\,\hat{u}_{x}\Delta\Psi(\hat{\vec{u}},t+\tau\vert\hat{\vec{v}},t)=\\
&&\nonumber\hspace{1cm}-\sum_{l=0}^{\infty}\sum_{m=-l}^{l}Y_{lm}^{*}(\hat{\vec{v}})\int_{t}^{t+\tau}ds\,\lambda(s)\int d\hat{\vec{u}}\sqrt{\frac{2\pi}{3}}\left\{\left[e^{-(2D_{r}-i\omega_{0})(t+\tau-s)}Y_{11}(\hat{\vec{u}})\right.\right.\\
&&\hspace{1cm}\left.\left.+e^{-(2D_{r}+i\omega_{0})(t+\tau-s)}Y_{1-1}(\hat{\vec{u}})\right]e^{-(l(l+1)D_{r}+im\omega_{0})s}{\cal L}_{1}Y_{lm}(\hat{\vec{u}})\right\}
\end{eqnarray}

After introducing these expressions into eq. (\ref{2.68}) we will obtain three terms corresponding to an expansion of the correlation function in powers of $\lambda(t)$, of zeroth, first and second order respectively. The presence of this driving yields an explicit dependence of the correlation function on the time $t$, instead of depending only on the time difference, as occurs in the stationary case. The method for removing this dependence on the initial time is to average the correlation function over a period of the driving \cite{kn:jung}. After doing this the first order term vanishes, and consequently we do not worry about it and compute only those whose average gives a non-zero contribution, \emph{i.e.} the zeroth and second order terms. Taking this into account, and by applying eq. (\ref{2.50}) to $\Psi(\hat{\vec{v}},t\vert\hat{\vec{R}}_{0},t_{0})$ 

\begin{eqnarray}\label{2.72}
\nonumber\langle\hat{R}_{x}(t)\hat{R}_{x}(t+\tau)\vert\hat{\vec{R}}_{0}(t_{0})\rangle&\sim&\int\,d\hat{\vec{v}}\int\,d\hat{\vec{u}}\,\hat{u}_{x}\hat{v}_{x}\Psi_{0}(\hat{\vec{v}},t\vert\hat{\vec{R}}_{0},t_{0})\Psi_{0}(\hat{\vec{u}},t+\tau\vert\hat{\vec{v}},t)\\
\nonumber&&+\int\,d\hat{\vec{v}}\int\,d\hat{\vec{u}}\,\hat{u}_{x}\hat{v}_{x}\Delta\Psi(\hat{\vec{v}},t\vert\hat{\vec{R}}_{0},t_{0})\Delta\Psi(\hat{\vec{u}},t+\tau\vert\hat{\vec{v}},t),\\
\mbox{}
\end{eqnarray}

\noindent where the sign $\sim$ indicates that the terms which vanish after averaging over the period of the driving has been neglected (although the average has not been performed yet). After introducing the corresponding expressions for $\Psi(\hat{\vec{v}},t\vert\hat{\vec{R}}_{0},t_{0})$ and by using eq. (\ref{2.71}) we obtain (the details of this computation are given in Appendix B)

\begin{equation}\label{2.73}
\langle\hat{R}_{x}(t)\hat{R}_{x}(t+\tau)\rangle\sim\left(\frac{4\pi}{3}\right)^{2}2e^{-2D_{r}\tau}\cos(\omega_{0}\tau)+\left(\frac{2\pi}{3}\right)^{3}\lambda^{2}(t)e^{i\omega\tau}\left\vert\int_{0}^{\infty}\,dt^{\prime}e^{i\omega t^{\prime}}\chi_{x}(t^{\prime})\right\vert^{2},
\end{equation}

\noindent where we have defined

\begin{equation}\label{2.74}
\langle\hat{R}_{x}(t)\hat{R}_{x}(t+\tau)\rangle\equiv\frac{1}{4\pi}\int\,d\hat{\vec{R}}_{0}\nonumber\langle\hat{R}_{x}(t)\hat{R}_{x}(t+\tau)\vert\hat{\vec{R}}_{0}(t_{0})\rangle.
\end{equation}

\noindent At this stage, and before applying the Fourier transform to the correlation function to obtain the power spectrum of the process $\hat{R}_{x}(t)$, we average eq. (\ref{2.73}) to obtain

\begin{eqnarray}\label{2.75}
\nonumber\overline{\langle\hat{R}_{x}(t)\hat{R}_{x}(t+\tau)\rangle}&=&\frac{\omega}{2\pi}\int_{0}^{\frac{2\pi}{\omega}}dt\,\langle\hat{R}_{x}(t)\hat{R}_{x}(t+\tau)\rangle\\
\nonumber &=&\left(\frac{4\pi}{3}\right)^{2}2e^{-2D_{r}\tau}\cos(\omega_{0}\tau)+\left(\frac{2\pi}{3}\right)^{3}\lambda_{0}^{2}e^{i\omega\tau}\left\vert\int_{0}^{\infty}\,dt^{\prime}e^{i\omega t^{\prime}}\chi_{x}(t^{\prime})\right\vert^{2},\\
\mbox{}
\end{eqnarray}

This computation has been carried out in the assumption that $\tau$ is a positive quantity. To extend our computation to $\tau<0$ we have to use the backwards Fokker-Planck equation. The operator which generates the backwards evolution of the probability distribution $-{\cal L}^{\dagger}$ \cite{kn:ris2} being ${\cal L}$ the Fokker-Planck operator and ${\cal L}^{\dagger}$ its adjoint operator. Consequently the formal solution of the backwards Fokker-Planck, equivalent to eq. (\ref{2.50}) is given by $(t<t_{0})$

\begin{equation}
\Psi(\hat{\vec{R}},t)\,=\,e^{-(t-t_{0}){\cal L}_{0}^{\dagger}}\Psi^{0}(t_{0})+
\int_{t_{0}}^{t}\,dt^{\prime}\,\lambda(t^{\prime})e^{-(t-t^{\prime}){\cal 
L}_{0}^{\dagger}}{\cal L}_{1}\Psi^{0}(t^{\prime}).
\end{equation}

\noindent As in the case of the operator ${\cal L}_{0}$, the spherical harmonics are eigenfunctions of $-{\cal L}_{0}^{\dagger}$ with eigenvalues given by

\begin{equation}
-{\cal L}_{0}^{\dagger}Y_{lm}(\hat{\vec{R}})=(l(l+1)D_{r}-im\omega_{0})Y_{lm}(\hat{\vec{R}}),
\end{equation}

\noindent Thus the process to follow in the calculation of the correlation function for $\tau<0$ is identical to the corresponding computation for $\tau>0$ changing the eigenvalues of the operator ${\cal L}_{0}$ by the ones of $-{\cal L}_{0}^{\dagger}$, which yields

\begin{eqnarray}
\nonumber\overline{\langle\hat{R}_{x}(t)\hat{R}_{x}(t+\tau)\rangle}=\left(\frac{4\pi}{3}\right)^{2}2e^{2D_{r}\tau}\cos(\omega_{0}\tau)+\left(\frac{2\pi}{3}\right)^{3}\lambda_{0}^{2}e^{i\omega\tau}\left\vert\int_{0}^{\infty}\,dt^{\prime}e^{i\omega t^{\prime}}\chi_{x}(t^{\prime})\right\vert^{2},\\
\mbox{}
\end{eqnarray} 

\noindent for $\tau<0$.

We now apply to this averaged correlation function (now defined for $-\infty<\tau<\infty$) the Wiener-Khinchine theorem, which states that the power spectrum and the correlation function are related through a Fourier transform. Thus, 

\begin{eqnarray}\label{2.76}
\nonumber Q(\Omega)&=&\int_{-\infty}^{\infty}d\tau\,\overline{\langle\hat{R}_{x}(t)\hat{R}_{x}(t+\tau)\rangle}e^{i\Omega\tau}=N(\Omega)+S(\omega)\delta(\Omega-\omega)\\
\nonumber N(\Omega)&=&\left(\frac{4\pi}{3}\right)^{2}\left[\frac{2D_{r}}{4D_{r}^{2}+(\Omega+\omega_{0})^{2}}+\frac{2D_{r}}{4D_{r}^{2}+(\Omega-\omega_{0})^{2}}\right],\\
S(\omega)&=&\left(\frac{2\pi}{3}\right)^{3}\lambda_{0}^{2}\,\vert\chi_{x}(\omega)\vert^{2}.
\end{eqnarray}

Since our purpose is to discern whether or not SR is present in the relaxation process of the quantity $\hat{R}_{x}(t)$, we proceed to compute the signal to noise ratio, $R$, \emph{i.e.} the ratio between the weight of the delta function in (\ref{2.76})$_{1}$ and the noisy part of $Q(\Omega)$ computed at the frequency of the driving. From eqs. (\ref{2.63}) and (\ref{2.76}) we achieve

\begin{equation}\label{2.77}
R=\frac{S(\omega)}{N(\omega)}=\lambda_{0}^{2}\frac{6}{\pi}\frac{\left(\frac{2D_{r}}{4D_{r}^{2}+(\omega+\omega_{0})^{2}}+\frac{2D_{r}}{4D_{r}^{2}+(\omega-\omega_{0})^{2}}\right)^{2}+\left(\frac{\omega+\omega_{0}}{4D_{r}^{2}+(\omega+\omega_{0})^{2}}+\frac{\omega-\omega_{0}}{4D_{r}^{2}+(\omega-\omega_{0})^{2}}\right)^{2}}{\frac{2D_{r}}{4D_{r}^{2}+(\omega+\omega_{0})^{2}}+\frac{2D_{r}}{4D_{r}^{2}+(\omega-\omega_{0})^{2}}}.
\end{equation}

\noindent This quantity has been plotted in Fig. 1 as a function of the inverse of the P\'eclet number, $Pe^{-1}=D_{r}/\omega_{0}$, which measures the ratio between the time scales associated to diffusion (thermal noise) and flow. The presence of a maximum in $R$ for a non-zero value of this parameter shows the existence of stochastic resonance in the relaxation process of a dipole in a shear flow. In addition to the slow relaxation to the single attractor of the dynamics, our model includes another effect, which hides, to some extent, the SR profile. To understand this, note that even though the signal is too weak, it nevertheless causes the position of the attractor of the dynamics to vary, and so the output will always have a nonzero component at the signal frequency. This fact causes the SNR to go to infinity in the zero noise limit \cite{kn:wiesenfeld1}.

\section{Mean first passage time}

In this section we study the behaviour of the escape time distribution and the mean first passage time of the magnetic dipole immersed in a shear flow. To this end, we have to account for the fixed point orientations of eq. (\ref{2.42}) in the case $\lambda_{0}<1$. In this situation there is a single fixed point corresponding to an orientation contained in the plane $x=0$ or, equivalently, $\phi=\pi/2$. However, when $\lambda(t)>0$ this stationary orientation is in the subspace $z>0$ ($\cos\theta>0$) and in the subspace $z<0$ ($\cos\theta<0$) if $\lambda(t)<0$. Therefore, we are going to study the escape from the region $z>0$ ($\cos\theta>0$) assuming that the initial orientation of the dipole is contained in this region. Consequently, we have to solve the Fokker-Planck equation (\ref{2.48}) with absorbing boundary conditions in the plane $\cos\theta=0$ \cite{kn:wei,kn:fhw}, \emph{i.e.}

\begin{equation}\label{2.81}
\Psi(\cos\theta=0,\phi,t)=0.
\end{equation}

Since this escape problem will be treated perturbatively, the first step is to analyze the eigenvalue problem of the operator ${\cal L}_{0}$ eq. (\ref{2.49}) under the boundary condition (\ref{2.81}). It is easy to check that the eigenfunctions and the eigenvalues are the same with the restriction that only those spherical harmonics which vanish at $\cos\theta=0$ are solution of this eigenvalue problem. From the parity properties of the associated Legendre functions \cite{kn:ch} one can see that (\ref{2.81}) selects only the spherical harmonics such that $l+m=2n+1$ with $n=0,1,...$. Thus, we have

\begin{equation}\label{2.82}
\Psi(\hat{\vec{R}},t)=\sum_{\langle l,m\rangle}a_{lm}(t)Y_{lm}(\hat{\vec{R}}),
\end{equation}

\noindent where $\langle l,m\rangle$ denotes that the sum is carried out over $0\leq l<\infty$ and $-l\leq m \leq l$ restricted by $l+m=2n+1$.

In order to evaluate the mean first passage time (MFPT), we have to compute previously the survival probability, $S(\hat{\vec{R}}_{0},t)$, and the escape time distribution (ETD) which are related through

\begin{equation}\label{2.83}
\rho(\hat{\vec{R}}_{0},t)=-\frac{d\,S(\hat{\vec{R}}_{0},t)}{dt},
\end{equation}

\noindent where $S(\hat{\vec{R}}_{0},t)$ is defined by

\begin{equation}\label{2.84}
S(\hat{\vec{R}}_{0},t)=\int_{{\cal R}}d\hat{\vec{R}}\,\Psi(\hat{\vec{R}},t\vert\hat{\vec{R}}_{0})=\int_{0}^{2\pi}d\phi\int_{0}^{1}d(\cos\theta)\Psi(\cos\theta,\phi,t\vert\hat{\vec{R}}_{0}),
\end{equation}

\noindent with ${\cal R}$, the region from which we are studying the escape problem (in the present case $\cos\theta>0$), and $\hat{\vec{R}}_{0}\in{\cal R}$, the initial orientation of the dipole. The probability distribution $\Psi(\hat{\vec{R}},t\vert\hat{\vec{R}}_{0})$ is obtained from eq. (\ref{2.50}) with the boundary conditions (\ref{2.81}) and the initial condition

\begin{equation}\label{2.85}
\Psi(\hat{\vec{R}},t=0)=\delta(\hat{\vec{R}}-\hat{\vec{R}}_{0})=\sum_{\langle l,m\rangle}Y_{lm}^{*}(\hat{\vec{R}}_{0})Y_{lm}(\hat{\vec{R}}).
\end{equation}

Before proceeding to obtain the survival probability, there are some facts to consider which will facilitate further computation. Looking at eq. (\ref{2.84}) one can realize that, due to the integration over the azimuthal angle, only terms with $m=0$ contribute to $S(\hat{\vec{R}}_{0},t)$. Consequently, the selection rule $l+m=2n+1$ reduces to keep only the odd values of $l$. In addition, we are interested only in the modes with greater relaxation times. Therefore, from the whole series eq. (\ref{2.82}) we are only interested in the term $l=1$, $m=0$. Thus, our purpose is to obtain the coefficient $a_{10}(t)$ up to first order in $\lambda(t)$ from eqs. (\ref{2.48}) and (\ref{2.81}). Up to zeroth order, we have

\begin{equation}\label{2.86}
a_{10}^{(0)}(t)=e^{-2D_{r}t}Y_{10}(\hat{\vec{R}}_{0}).
\end{equation}

The obtention of the first order contribution, $a_{10}^{(1)}(t)$, requires a little bit more elaborated calculation. To proceed further with this computation, the operator ${\cal L}_{1}$ acting on $Y_{lm}(\hat{\vec{R}})$ yields

\begin{eqnarray}\label{2.87}
\nonumber{\cal L}_{1}Y_{lm}(\hat{\vec{R}})&=&-2\omega_{0}\sqrt{\frac{2\pi}{3}}(Y_{11}(\hat{\vec{R}})+Y_{1-1}(\hat{\vec{R}}))Y_{lm}(\hat{\vec{R}})\\
&-&\omega_{0}\left[\sqrt{\frac{4\pi}{3}}Y_{10}(\hat{\vec{R}}){\cal R}_{y}-i\sqrt{\frac{2\pi}{3}}(Y_{11}(\hat{\vec{R}})-Y_{1-1}(\hat{\vec{R}})){\cal R}_{z}\right]Y_{lm}(\hat{\vec{R}}),
\end{eqnarray}

\noindent where the action of ${\cal R}_{z}$ on $Y_{lm}(\hat{\vec{R}})$ is given by eq. (\ref{2.52})$_{1}$ and

\begin{equation}\label{2.88}
{\cal R}_{y}Y_{lm}(\hat{\vec{R}})=-\frac{1}{2}\left\{\sqrt{(l-m)(l+m+1)}Y_{lm+1}(\hat{\vec{R}})-\sqrt{(l+m)(l-m+1)}Y_{lm-1}(\hat{\vec{R}})\right\}.
\end{equation}

\noindent From eqs. (\ref{2.87}) and (\ref{2.88}) together with the rules for the addition of angular momenta familiar from quantum mechanics \cite{kn:sak} and the selection rule $l+m=2n+1$ imposed by the boundary condition (\ref{2.81}), one can deduce that only the terms $Y_{2\pm 1}(\hat{\vec{R}})$ contributes to $a_{10}^{(1)}(t)$. The mentioned rules of addition of angular momenta implies that the product of two spherical harmonics, $Y_{lm}Y_{pq}$ has projection onto a third spherical harmonic, $Y_{rs}$, only when $m+q=s$. On the other hand these same rules imposes the restriction that the product $Y_{lm}Y_{pq}$ only projects onto the subspaces such that $\vert l-p\vert\leq r\leq l+p$. By using these restrictions one can see that when in eq. (\ref{2.87}) one takes $l=1$ and $m=0$ one obtains a vanishing contribution and only when $l=2$ and $m=\pm 1$ the contribution to $a_{10}^{(1)}(t)$ is different from zero. All other contribution of higher values $l$ are explicitly excluded by the rule $\vert l-p\vert\leq r\leq l+p$. 

Taking these considerations into account and by using the following results 

\begin{eqnarray}\label{2.89}
\nonumber&&\frac{3}{4\pi}\int_{0}^{2\pi}d\phi\int_{0}^{1}d(\cos\theta)Y_{11}(\hat{\vec{R}})Y_{2-1}(\hat{\vec{R}})Y_{10}(\hat{\vec{R}})=\frac{9}{80\pi}\sqrt{\frac{5}{3\pi}},\\
\nonumber&&\frac{3}{4\pi}\int_{0}^{2\pi}d\phi\int_{0}^{1}d(\cos\theta)Y_{1-1}(\hat{\vec{R}})Y_{21}(\hat{\vec{R}})Y_{10}(\hat{\vec{R}})=\frac{9}{80\pi}\sqrt{\frac{5}{3\pi}},\\
&&\frac{3}{4\pi}\int_{0}^{2\pi}d\phi\int_{0}^{1}d(\cos\theta)Y_{10}(\hat{\vec{R}})Y_{20}(\hat{\vec{R}})Y_{10}(\hat{\vec{R}})=\frac{3}{20\pi}\sqrt{\frac{5}{4\pi}}.
\end{eqnarray}

\noindent we obtain the first order correction to the coefficient $a_{10}(t)$.

\begin{eqnarray}\label{2.90}
\nonumber&& a_{10}^{(1)}(t)=\\
\nonumber&&-\omega_{0}\frac{\lambda_{0}}{40}\sqrt{\frac{15}{4\pi}}\left(\frac{\sqrt{2}}{2}+2\sqrt{3}\right)\left\{Y_{21}(\hat{\vec{R}}_{0})\frac{4D_{r}+i(\omega-\omega_{0})}{16D_{r}^{2}+(\omega-\omega_{0})^{2}}+ Y_{2-1}(\hat{\vec{R}}_{0})\frac{4D_{r}+i(\omega+\omega_{0})}{16D_{r}^{2}+(\omega+\omega_{0})^{2}}\right\}.\\
\mbox{}
\end{eqnarray}

\noindent Eqs. (\ref{2.86}) and (\ref{2.90}) together with (\ref{2.82}) and (\ref{2.84}) allows us to obtain the survival probability $S(\hat{\vec{R}}_{0},t)$, which is given by

\begin{eqnarray}\label{2.90b}
\nonumber S(\hat{\vec{R}}_{0},t)&=&\pi\sqrt{\frac{3}{4\pi}}Y_{10}(\hat{\vec{R}}_{0})e^{-2D_{r}t}\\
\nonumber &-&\frac{\omega_{0}}{40}\sqrt{\frac{15}{4\pi}}\left(\frac{\sqrt{2}}{2}+2\sqrt{3}\right)\left\{Y_{21}(\hat{\vec{R}}_{0})e^{-2D_{r}t}\int_{0}^{t}d\tau\lambda(\tau)e^{-(4D_{r}+i\omega_{0})\tau}\right.\\
&&\left.+Y_{2-1}(\hat{\vec{R}}_{0})e^{-2D_{r}t}\int_{0}^{t}d\tau\lambda(\tau)e^{-(4D_{r}-i\omega_{0})\tau}\right\}
\end{eqnarray}

This quantity is directly related to the MFPT, since

\begin{equation}\label{2.91}
T(\hat{\vec{R}}_{0})=\int_{0}^{\infty}dt\,t\,\rho(\hat{\vec{R}}_{0},t)=\int_{0}^{\infty}dt\,S(\hat{\vec{R}}_{0},t),
\end{equation}

\noindent where we have used eq. (\ref{2.83}). Consequently, the MFPT is given by

\begin{eqnarray}\label{2.92}
\nonumber T(\hat{\vec{R}}_{0})&\equiv& T_{0}(\hat{\vec{R}}_{0})+\Delta T(\hat{\vec{R}}_{0})\\
\nonumber T_{0}(\hat{\vec{R}}_{0})&=&\pi\sqrt{\frac{3}{4\pi}}\frac{\sqrt{1-\lambda^{2}}}{2D_{r}},\\
\nonumber\Delta T(\hat{\vec{R}}_{0})&=&T_{0}\left[\frac{\lambda_{0}^{2}}{40}\sqrt{\frac{15}{4\pi}}\left(\frac{\sqrt{2}}{2}+2\sqrt{3}\right)3\omega_{0}\sqrt{\frac{5}{24\pi}}\left\{\frac{(\omega-\omega_{0})}{16D_{r}^{2}+(\omega-\omega_{0})^{2}}+ \frac{(\omega+\omega_{0})}{16D_{r}^{2}+(\omega+\omega_{0})^{2}}\right\}\right],\\
\mbox{}
\end{eqnarray}

\noindent where we have taken $\hat{\vec{R}}_{0}=\hat{\vec{R}}_{s}$. In Fig. 2 we have plotted the quantity $\Delta T/T_{0}$. This figure shows that this quantity exhibits a minimum, as required for the appearance of SR.

The knowledge of the survival probability allows us to obtain the ETD, $\rho(\hat{\vec{R}}_{0},t)$. From eqs. (\ref{2.83}) and (\ref{2.90b}) the ETD is given by

\begin{eqnarray}\label{2.94}
\nonumber\rho(\hat{\vec{R}}_{0},t)&=&e^{-2D_{r}t}\left[2D_{r}\pi\sqrt{\frac{3}{4\pi}}Y_{10}(\hat{\vec{R}}_{0})-\lambda_{0}\frac{2D_{r}}{40}\sqrt{\frac{15}{4\pi}}\left(\frac{\sqrt{2}}{2}+2\sqrt{3}\right)\left\{Y_{21}(\hat{\vec{R}}_{0})\frac{2D_{r}+i(\omega-\omega_{0})}{16D_{r}^{2}+(\omega-\omega_{0})^{2}}\right.\right.\\
\nonumber&&\left.\left.Y_{2-1}(\hat{\vec{R}}_{0})\frac{2D_{r}+i(\omega+\omega_{0})}{16D_{r}^{2}+(\omega+\omega_{0})^{2}}\right\}+\lambda_{0}\frac{1}{40}\sqrt{\frac{15}{4\pi}}\left(\frac{\sqrt{2}}{2}+2\sqrt{3}\right)\left\{Y_{21}(\hat{\vec{R}}_{0})e^{-(4D_{r}+i(\omega-\omega_{0}))t}\right.\right.\\
&&\left.\left.Y_{2-1}(\hat{\vec{R}}_{0})e^{-(4D_{r}+i(\omega+\omega_{0}))t}\right\}\right]
\end{eqnarray}

\noindent By taking $\hat{\vec{R}}_{0}=\hat{\vec{R}}_{s}$, we finally obtain

\begin{eqnarray}\label{2.95}
\nonumber\frac{e^{2D_{r}t}}{2D_{r}\sqrt{1-\lambda_{0}^{2}}}\Delta\rho&=&\frac{\omega_{0}\lambda_{0}^{2}}{40\pi}\sqrt{75}\left(\frac{\sqrt{2}}{2}+2\sqrt{3}\right)\left\{\frac{e^{-4D_{r}t}}{2}(\sin[(\omega-\omega_{0})t]+\sin[(\omega+\omega_{0})t])\right.\\
&&\left.-\left[\frac{(\omega-\omega_{0})}{16D_{r}^{2}+(\omega-\omega_{0})^{2}}+ \frac{(\omega+\omega_{0})}{16D_{r}^{2}+(\omega+\omega_{0})^{2}}\right]\right\},
\end{eqnarray}

\noindent where $\Delta\rho\equiv\rho-\rho_{0}$, being $\rho_{0}$ the corresponding ETD when the amplitude of the oscillating field is set to zero.

The succession of maxima in the ETD (see Fig. 3) is indicating that the dynamic of the magnetic dipole suspended in a shear flow under a periodic field exhibits SR. 

\section{Discussion}

%We have shown that, unlike a dipole in a fluid at rest \cite{kn:vilar} undergoing Debye relaxation, a dipole in shear flow exhibits SR for low intensity magnetic field. The reason for this behaviour is that our system relaxes onto the attractor of the dynamics in a slower way. That is why we can observe a non-monotonous SNR in the linear regime. We have also computed the ETD, and the MFPT, both quantifiers of the SR phenomenon too. The ETD shows an oscillatory behavior and the MFPT a minimum as function of the frequency, respectively. Thus we can conclude that this feature of the relaxation process described in this paper, is a peculiarity of the non-equilibrium nature of the system.

We have shown that the relaxation process of a dipole immersed in a shear flow exhibits SR upon application of a weak periodic field. To this end we have computed three quantities typically used to characterize SR, namely the signal-to-noise ratio, the escape-time distribution and the mean first passage time. All of them behave as expected for a process in which SR occurs. 

Previous works devoted to analyze whether or not SR is present in the relaxation process of an overdamped dipole in a fluid at rest have revealed that this phenomenon does not occur in the linear regime \cite{kn:vilar}. Effectively, linear response theory (LRT) predicts a maximum in the signal, \emph{i.e.} in the susceptibility, as a function of the noise level. Unlikely, the SNR decreases monotonically with the noise level. This behaviour can be easily understood. In the limit of zero noise the output of the system has a small component (proportional to the applied field) at the frequency of the signal whereas the background noise vanishes at zero noise level, being this behaviour the responsible for the monotonic dependence of the SNR on the noise intensity. 

In our case, the situation is completely different. When the fluid in which the dipole is suspended is submitted to a pure rotation (vortex flow), both output signal and output background noise exhibit a peak at the same value of $D_{r}$ (see Fig. 4). Consequently, although the background noise vanishes when $D_{r}$ goes to zero, the characteristic SR profile of the SNR can not be completely hidden, as shown in Fig. 1. This feature arises as a consequence of the presence of shear acting on the suspension, thus, the appearance of SR in the system studied in this paper is a non-equilibrium feature. 

In a sense, the mechanism yielding SR in this system is similar to the one operating in SR in threshold devices \cite{kn:kiss,kn:cbg}. Due to the presence of noise, the dipole can eventually acquire enough energy to get out from its stable orientation by crossing the absorbing barrier (the threshold) $\cos\theta=0$. After this, the system is driven to its stable position. This process produces a short spike in the magnetization. Of course, the time that the system takes to return to the fixed point has to be smaller than the semiperiod of the oscillating magnetic field. Thus, SR in this system can be understood in the same way as, for example, the SR in level crossing detectors \cite{kn:kiss}.

LRT has been one of the widest used tools in the study of stochastic resonance \cite{kn:dykman1}. When the system, in the absence of the external periodic force, is in thermal equilibrium a very adequate form of describing stochastic resonance is in terms of the susceptibility. This is because the noisy part of the power spectrum is given directly by the susceptibility through the fluctuation-dissipation theorem,

\begin{equation}\label{2.100}
{\sf Im}\chi(\Omega)=\frac{\Omega}{2D_{r}}N(\Omega),
\end{equation} 

\noindent This result is correct when the fluctuations whose spectral density is given by $N(\Omega)$ have the thermal equilibrium state as reference state \cite{kn:rdl}.

However, in the present case we are dealing with a system which is maintained in an out-of-equilibrium steady state due to the presence of a shear flow. It is evident from Fig. 5, where we have plotted the imaginary part of the susceptibility corresponding to $\hat{R}_{x}$ and the noisy part of the power spectrum, that these two quantities are clearly different. Note, that if $\omega_{0}=0$, \emph{i.e.} the system in the absence of the periodic field is in equilibrium, the relation (\ref{2.100}) is fulfilled. Thus, we have shown that although we can define a susceptibility which describes the response of our system to a small perturbation we can not describe SR by means of LRT. The reason can be found in the fact that due to the non-equilibrium nature of the attractor of the dynamics the fluctuation-dissipation theorem fails to be valid.

\begin{center}{\bf ACKNOWLEDGMENTS}\end{center}

The authors thank Miguel Rub\'{\i} for valuable discussions. This work has been supported by DGICYT of the Spanish Government under grant PB98-1258. One of us (T. Alarc\'on) wishes to thank to DGICYT of the Spanish Government for financial support.

\section*{Appendix A: linear stability analysis of the fixed points of eq. (\ref{2.42})}

\addcontentsline{toc}{section}{Appendix A: linear stability analysis of the fixed points of eq. (\ref{2.42})}
\setcounter{equation}{0}
\def\theequation{A.\arabic{equation}}
\def\thesection{A.\arabic{section}}
\def\thesubsection{\thesection.\arabic{subsection}}
\setcounter{section}{0}
\setcounter{subsection}{0}
\setcounter{figure}{0}
\def\thefigure{A.\arabic{figure}}

From eqs. (\ref{2.41}) and (\ref{2.42}) one can see that the time derivative of the $\hat{\vec{R}}$ vanishes either when $\vec{\Omega}_{p}=0$ or when $\vec{\Omega}_{p}\times\hat{\vec{R}}=0$. In the first case we have 

\begin{equation}\label{a:1}
\vec{\Omega}_{p}=\omega_0\left\{\hat{\vec{z}} \right.+ 
\lambda_{0}
\left.\left (\hat{\vec{R}}_{s}\times\hat{\vec{x}}\right )\right\}=0\;\Rightarrow\;\hat{\vec{z}}=-\lambda_{0}\hat{\vec{R}}_{s}\times\hat{\vec{x}}.
\end{equation}

\noindent From this equation, and taking into account that $\vert\hat{\vec{R}}_{s}\vert=1$, we obtain that the stationary orientation is

\begin{equation}\label{a:2}
\hat{\vec{R}}_{s}=\pm\sqrt{1-\lambda_{0}^{-2}}\hat{\vec{x}}+\lambda_{0}^{-1}\hat{\vec{y}}.
\end{equation}

\noindent This solution exists only when $\lambda_{0}\geq 1$ and corresponds to a fixed orientation of the dipoles, given that the intensity of the magnetic field is high enough to maintain this fixed direction.

The second possibility leads to 

\begin{equation}\label{a:3}
\hat{\vec{z}}\times\hat{\vec{R}}_{s}=-\lambda_{0}(\hat{\vec{R}}_{s}\times\hat{\vec{x}})\times\hat{\vec{R}}_{s}=-\lambda_{0}(\hat{\vec{x}}-(\hat{\vec{x}}\cdot\hat{\vec{R}}_{s})\hat{\vec{R}}_{s}).
\end{equation}

\noindent Eq. (\ref{a:3}) provides two equations for three unknowns. If one sets $\hat{R}_{z}=0$ one recovers eq. (\ref{a:2}). If, by contrast one makes $\hat{R}_{x}=0$ then a new stationary orientation is obtained,

\begin{equation}\label{a:4}
\hat{\vec{R}}_{s}=\lambda_{0}\hat{\vec{y}}\pm\sqrt{1-\lambda_{0}^{2}}\hat{\vec{z}},
\end{equation}

\noindent which exists only when $\lambda_{0}\leq 1$. This orientation gives rise to a rotation of the dipoles with angular velocity

\begin{equation}\label{a:5}
\vec{\Omega}_{s}=\omega_{0}\sqrt{1-\lambda_{0}^{2}}\left\{\sqrt{1-\lambda_{0}^{2}}\hat{\vec{z}}\pm\lambda_{0}\hat{\vec{y}}\right\},
\end{equation}

\noindent since, in this case, the field is not strong enough to inhibit the rotation caused by the shear flow.

The linear stability of this fixed points is better analyzed in spherical coordinates. Taking into account that

\begin{equation}\label{a.1}
(\hat{\vec{R}}\times\hat{\vec{x}})\times\hat{\vec{R}}=\hat{\vec{x}}-\hat{\vec{R}}(\hat{\vec{R}}\cdot\hat{\vec{x}}),
\end{equation}

\noindent we obtain

\begin{eqnarray}\label{a.2}
\nonumber\frac{1}{\omega_{0}}\frac{d\hat{R}_{x}}{dt}&=&\lambda(1-\hat{R}_{x}^{2})-\hat{R}_{y},\\
\nonumber\frac{1}{\omega_{0}}\frac{d\hat{R}_{y}}{dt}&=&-\lambda\hat{R}_{x}\hat{R}_{y}+\hat{R}_{x},\\
\frac{1}{\omega_{0}}\frac{d\hat{R}_{z}}{dt}&=&-\lambda\hat{R}_{x}\hat{R}_{z}.
\end{eqnarray}

\noindent After expressing the components of $\hat{\vec{R}}$ in spherical coordinates, we obtain the following bidimensional dynamical system

\begin{eqnarray}\label{a.3}
\nonumber\frac{1}{\omega_{0}}\frac{d\theta}{dt}&=&\lambda\cos\theta\cos\phi,\\
\frac{1}{\omega_{0}}\frac{d\phi}{dt}&=&-\lambda\frac{\sin\phi}{\sin\theta}+1.
\end{eqnarray}

\noindent where $\theta$ and $\phi$ are the polar and azimuthal angles, respectively. By linearization of eqs. (\ref{a.3}) around the $\lambda\geq 1$ fixed points we obtain the following matrix 

\begin{equation}\label{a.8} 
{\sf A}(\lambda\geq 1)=\left( \begin{array}{cc} \mp\lambda\sqrt{1-\left(\frac{1}{\lambda}\right)^{2}} & 0 \\ 0 &  \mp\lambda\sqrt{1-\left(\frac{1}{\lambda}\right)^{2}} \end{array} \right) \end{equation}

\noindent which implies that, if $\lambda$ is positive, the orientation corresponding to choose the sign "+" in eq. (\ref{a:2}) is stable, while the other one is unstable.

The same linearization procedure carried out around the $\lambda< 1$ fixed points leads to 

\begin{equation}\label{a.9} 
{\sf A}(\lambda<1)=\left( \begin{array}{cc} 0 & \lambda^{2}\sqrt{1-\lambda^{2}} \\ -\lambda^{2}\sqrt{1-\lambda^{2}} & 0 \end{array} \right) \end{equation}

\noindent The eigenvalues of this matrix are given by

\begin{equation}\label{a.10}
\alpha=\pm i\lambda^{2}\sqrt{1-\lambda^{2}}.
\end{equation}

\section*{Appendix B: computation of eqs. (\ref{2.71}) and (\ref{2.73}).}

\addcontentsline{toc}{section}{Appendix B: computation of eqs. (\ref{2.71}) and (\ref{2.73}).}
\setcounter{equation}{0}
\def\theequation{B.\arabic{equation}}
\def\thesection{B.\arabic{section}}
\def\thesubsection{\thesection.\arabic{subsection}}
\setcounter{section}{0}
\setcounter{subsection}{0}
\setcounter{figure}{0}
\def\thefigure{B.\arabic{figure}}

In this appendix we work in detail some steps of the computation of the power spectrum corresponding to the relaxation process of a dipole under an oscillating magnetic field in a shear flow, in particular we calculate the integrals which yield eqs. (\ref{2.71}) and (\ref{2.73}). From eqs. (\ref{2.69}) and (\ref{2.70})

\begin{eqnarray}\label{b.1}
\nonumber&&\int\,d\hat{\vec{u}}\,\hat{u}_{x}\Psi(\hat{\vec{u}},t+\tau\vert\hat{\vec{v}},t)=I_{1}+I_{2},\\
\nonumber &&I_{1}=\int\,d\hat{\vec{u}}\,\hat{u}_{x}\Psi_{0}(\hat{\vec{u}},t+\tau\vert\hat{\vec{v}},t)=\int\,d\hat{\vec{u}}\,\hat{u}_{x}e^{\tau{\cal L}_{0}}\delta(\hat{\vec{u}}-\hat{\vec{v}}),\\
\nonumber &&I_{2}=\int\,d\hat{\vec{u}}\,\hat{u}_{x}\Delta\Psi(\hat{\vec{u}},t+\tau\vert\hat{\vec{v}},t)=\int\,d\hat{\vec{u}}\,\hat{u}_{x}\int_{t}^{t+\tau}ds\,\lambda(s)e^{(t+\tau-s){\cal L}_{0}}{\cal L}_{1}e^{s{\cal L}_{0}}\delta(\hat{\vec{u}}-\hat{\vec{v}}).\\
\mbox{}
\end{eqnarray}

To begin with we focus on the integral $I_{1}$, which can be rewritten as

\begin{equation}\label{b.2}
I_{1}=\int\,d\hat{\vec{u}}\,\left(e^{\tau{\cal L}_{0}^{\dagger}}\hat{u}_{x}\right)\delta(\hat{\vec{u}}-\hat{\vec{v}}),
\end{equation}

\noindent where ${\cal L}_{0}^{\dagger}$ is the adjoint operator of ${\cal L}_{0}$ defined by 

\begin{equation}\label{b.3}
\int\,d\hat{\vec{u}}\,{\sf A}\,({\cal L}_{0}{\sf B})=\int\,d\hat{\vec{u}}\,({\cal L}_{0}^{\dagger}{\sf A})\,{\sf B},
\end{equation}

\noindent with ${\sf A}$ and ${\sf B}$ two arbitrary observables. Explicitly, ${\cal L}_{0}^{\dagger}$ is given by

\begin{eqnarray}\label{b.4}
\nonumber&&{\cal L}_{0}^{\dagger}=\omega_{0}\hat{\vec{R}}_{0}\cdot\vec{{\cal R}}+D_{r}{\cal R}^{2},\\
&&{\cal L}_{0}^{\dagger}Y_{lm}(\hat{\vec{R}})=(-l(l+1)D_{r}+i\omega_{0}m)Y_{lm}(\hat{\vec{R}}),
\end{eqnarray}

\noindent and therefore eq. (\ref{b.2}) reads

\begin{eqnarray}\label{b.5}
\nonumber I_{1}&=&-\sqrt{\frac{2\pi}{3}}\int\,d\hat{\vec{u}}\left[e^{-(2D_{r}-i\omega_{0})\tau}Y_{11}(\hat{\vec{u}})+e^{-(2D_{r}+i\omega_{0})\tau}Y_{1-1}(\hat{\vec{u}})\right]\delta(\hat{\vec{u}}-\hat{\vec{v}})\\
&=&-\sqrt{\frac{2\pi}{3}}\left[e^{-(2D_{r}-i\omega_{0})\tau}Y_{11}(\hat{\vec{v}})+e^{-(2D_{r}+i\omega_{0})\tau}Y_{1-1}(\hat{\vec{v}})\right],
\end{eqnarray}

\noindent leading to eq. (\ref{2.71})$_{1}$. In eq. (\ref{b.5}) we have used the relation

\begin{equation}\label{b.6}
\hat{u}_{x}=-\sqrt{\frac{2\pi}{3}}(Y_{11}(\hat{\vec{u}})+Y_{1-1}(\hat{\vec{u}})).
\end{equation}

To compute the integral $I_{2}$, we have to use the following representation of the delta function

\begin{equation}\label{b.7}
\delta(\hat{\vec{u}}-\hat{\vec{v}})=\sum_{l=0}^{\infty}\sum_{m=-l}^{l}Y_{lm}^{*}(\hat{\vec{v}})Y_{lm}(\hat{\vec{u}}).
\end{equation}

\noindent After introducing this expression into eq. (\ref{b.1})$_{3}$ we obtain

\begin{eqnarray}\label{b.8}
\nonumber I_{2}&=&\sum_{l=0}^{\infty}\sum_{m=-l}^{l}Y_{lm}^{*}(\hat{\vec{v}})\int_{t}^{t+\tau}ds\,\lambda(s)\int\,d\hat{\vec{u}}\left(e^{(t+\tau-s){\cal L}_{0}^{\dagger}}\hat{u}_{x}\right){\cal L}_{1}e^{s{\cal L}_{0}}Y_{lm}(\hat{\vec{u}}),\\
\nonumber &=&\sum_{l=0}^{\infty}\sum_{m=-l}^{l}Y_{lm}^{*}(\hat{\vec{v}})\int_{t}^{t+\tau}ds\,\lambda(s)e^{-(l(l+1)D_{r}+im\omega_{0})s}\int\,d\hat{\vec{u}}\left(e^{(t+\tau-s){\cal L}_{0}^{\dagger}}\hat{u}_{x}\right){\cal L}_{1}Y_{lm}(\hat{\vec{u}}),\\
\mbox{}
\end{eqnarray}

\noindent and by using eq. (\ref{b.6}), eq. (\ref{b.8}) yields eq. (\ref{2.71}), \emph{i.e.}

\begin{eqnarray}\label{b.9}
\nonumber
I_{2}&=&-\sqrt{\frac{2\pi}{3}}\sum_{l=0}^{\infty}\sum_{m=-l}^{l}Y_{lm}^{*}(\hat{\vec{v}})\int_{t}^{t+\tau}ds\,\lambda(s)\int\,d\hat{\vec{u}}\left\{\left[e^{-(2D_{r}-i\omega_{0})(t+\tau-s)}Y_{11}(\hat{\vec{u}})\right.\right.\\
&&\left.\left.+e^{-(2D_{r}+i\omega_{0})(t+\tau-s)}Y_{1-1}(\hat{\vec{u}})\right]
e^{-(l(l+1)D_{r}+im\omega_{0})s}{\cal L}_{1}Y_{lm}(\hat{\vec{u}})\right\}
\end{eqnarray}

Once these expressions have been obtained we can compute the correlation function given by eq. (\ref{2.72}),

\begin{equation}\label{b.10}
\langle\hat{R}_{x}(t)\hat{R}_{x}(t+\tau)\vert\hat{\vec{R}}_{0}(t_{0})\rangle\sim\int\,d\hat{\vec{v}}\hat{v}_{x}\Psi_{0}(\hat{\vec{v}},t\vert\hat{\vec{R}}_{0},t_{0})I_{1}+
\int\,d\hat{\vec{v}}\hat{v}_{x}\Delta\Psi(\hat{\vec{v}},t\vert\hat{\vec{R}}_{0},t_{0})I_{2},
\end{equation}

\noindent being $\Psi_{0}(\hat{\vec{v}},t\vert\hat{\vec{R}}_{0},t_{0})$ and $\Delta\Psi(\hat{\vec{v}},t\vert\hat{\vec{R}}_{0},t_{0})$ given by

\begin{eqnarray}\label{b.11}
\nonumber\Psi_{0}(\hat{\vec{v}},t\vert\hat{\vec{R}}_{0})&=&e^{t{\cal L}_{0}}\delta(\hat{\vec{v}}-\hat{\vec{R}}_{0})=\sum_{l=0}^{\infty}\sum_{m=-l}^{l}Y_{lm}^{*}(\hat{\vec{R}}_{0})e^{-(l(l+1)D_{r}+im\omega_{0})t}Y_{lm}(\hat{\vec{v}}),\\
\nonumber\Delta\Psi(\hat{\vec{v}},t\vert\hat{\vec{R}}_{0})&=&\int_{0}^{t}dr\,\lambda(r)e^{(t-r){\cal L}_{0}}{\cal L}_{1}e^{r{\cal L}_{0}}\delta(\hat{\vec{v}}-\hat{\vec{R}}_{0})\\
\nonumber&=&\sum_{l=0}^{\infty}\sum_{m=-l}^{l}Y_{lm}^{*}(\hat{\vec{R}}_{0})\int_{0}^{t}dr\,\lambda(r)e^{-(l(l+1)D_{r}+im\omega_{0})r}e^{(t-r){\cal L}_{0}}{\cal L}_{1}Y_{lm}(\hat{\vec{v}}).\\
\mbox{}
\end{eqnarray}

\noindent where the initial time $t_{0}$ has been fixed to zero and eq. (\ref{b.7}) has been used. 

Eqs. (\ref{b.11}) provide the evolution of the probability distribution under the condition of the system being initially in the state $\hat{\vec{R}}_{0}$. Since \emph{a priori} nothing is known about this initial condition, we assume that $\hat{\vec{R}}_{0}$ is a random variable uniformly distributed over the orientation space, consequently we average the correlation function over the distribution of initial states (eq. (\ref{2.74})). Taking into account that

\begin{equation}\label{b.12}
\frac{1}{4\pi}\int\,d\hat{\vec{R}}_{0}\,Y_{lm}^{*}(\hat{\vec{R}}_{0})=\delta_{l,0}\delta_{m,0},
\end{equation}

\noindent the average of the correlation function over initial conditions

\begin{eqnarray}\label{b.13}
\nonumber&&\langle\hat{R}_{x}(t)\hat{R}_{x}(t+\tau)\rangle\sim I_{3}+I_{4}\\
\nonumber&& I_{3}\equiv\frac{1}{4\pi}\int\,d\hat{\vec{v}}\hat{v}_{x}I_{1},\\ 
&&I_{4}\equiv\frac{1}{\sqrt{4\pi}}\int\,d\hat{\vec{v}}\hat{v}_{x}\int_{0}^{t}dr\,\lambda(r)e^{(t-r){\cal L}_{0}}{\cal L}_{1}Y_{00}(\hat{\vec{v}})I_{2}.
\end{eqnarray}

\noindent After introducing eqs. (\ref{b.5}) and (\ref{b.6}) into (\ref{b.13})$_{2}$, we obtain

\begin{eqnarray}\label{b.14}
\nonumber I_{3}&=&\frac{1}{4\pi}\int\,d\hat{\vec{v}}\frac{2\pi}{3}(Y_{11}(\hat{\vec{v}})+Y_{1-1}(\hat{\vec{v}}))\left[e^{-(2D_{r}-i\omega_{0})\tau}Y_{11}(\hat{\vec{v}})+e^{-(2D_{r}+i\omega_{0})\tau}Y_{1-1}(\hat{\vec{v}})\right]\\
&=&\frac{4\pi}{9}e^{-2D_{r}\tau}\cos(\omega_{0}\tau),
\end{eqnarray}

\noindent where the orthogonality relation for the spherical harmonics,

\begin{equation}\label{b.15}
\int\,d\hat{\vec{v}}\,Y_{pq}^{*}(\hat{\vec{v}})Y_{lm}(\hat{\vec{v}})=\frac{4\pi}{2l+1}\frac{(l+m)!}{(l-m)!}\,\delta_{l,p}\delta_{m,q},
\end{equation}

\noindent has been used. On the other hand from eqs. (\ref{b.9}) and (\ref{b.13})$_{3}$ 

\begin{eqnarray}\label{b.16}
\nonumber\hspace{-1cm} I_{4}&=&-\frac{1}{3}\sqrt{\frac{2\pi}{3}}\sum_{l=0}^{\infty}\sum_{m=-l}^{l}\int_{0}^{t}dr\,\lambda(r)\int_{t}^{t+\tau}ds\,\lambda(s)\int\,d\hat{\vec{v}}\left\{(Y_{11}(\hat{\vec{v}})+Y_{1-1}(\hat{\vec{v}}))\left(e^{-(2D_{r}-i\omega_{0})(t-r)}Y_{11}(\hat{\vec{v}})\right.\right.\\ \nonumber&&\hspace{-0.7cm}\left.\left.+\;e^{-(2D_{r}+i\omega_{0})(t-r)}Y_{1-1}(\hat{\vec{v}})\right)Y_{lm}^{*}(\hat{\vec{v}})\int\,d\hat{\vec{u}}\left\{\left[e^{-(2D_{r}-i\omega_{0})(t+\tau-s)}Y_{11}(\hat{\vec{u}})+e^{-(2D_{r}+i\omega_{0})(t+\tau-s)}Y_{1-1}(\hat{\vec{u}})\right]\right.\right.\\
&&\hspace{-0.3cm}\left.\left.e^{-(l(l+1)D_{r}+im\omega_{0})s}{\cal L}_{1}Y_{lm}(\hat{\vec{u}})\right\}\right\}.
\end{eqnarray}

\noindent Let us focus our attention on the integral over $\hat{\vec{v}}$. 

\begin{eqnarray}\label{b.17a}
\nonumber&&\int\,d\hat{\vec{v}}(Y_{11}(\hat{\vec{v}})+Y_{1-1}(\hat{\vec{v}}))\left(e^{-(2D_{r}-i\omega_{0})(t-r)}Y_{11}(\hat{\vec{v}})+e^{-(2D_{r}+i\omega_{0})(t-r)}Y_{1-1}(\hat{\vec{v}})\right)Y_{lm}^{*}(\hat{\vec{v}})=\\
\nonumber&&\int\,d\hat{\vec{v}}Y_{11}(\hat{\vec{v}})e^{-(2D_{r}-i\omega_{0})(t-r)}Y_{11}Y_{lm}^{*}(\hat{\vec{v}})+\int\,d\hat{\vec{v}}Y_{1-1}(\hat{\vec{v}})e^{-(2D_{r}+i\omega_{0})(t-r)}Y_{1-1}(\hat{\vec{v}})Y_{lm}^{*}(\hat{\vec{v}})+\\
\nonumber&&\int\,d\hat{\vec{v}}Y_{1-1}(\hat{\vec{v}})e^{-(2D_{r}-i\omega_{0})(t-r)}Y_{11}(\hat{\vec{v}})Y_{lm}^{*}(\hat{\vec{v}})+\int\,d\hat{\vec{v}}Y_{11}(\hat{\vec{v}})e^{-(2D_{r}+i\omega_{0})(t-r)}Y_{1-1}(\hat{\vec{v}})Y_{lm}^{*}(\hat{\vec{v}}).\\
\mbox{}
\end{eqnarray}

From the rules of addition of angular momentum familiar from quantum mechanics, which imply that the product of two spherical harmonics, $Y_{pq}Y_{rs}$, only has a non-vanishing projection over a third spherical harmonic, $Y_{lm}$, when the relations $\vert p-r\vert\leq l\leq r+p$ and $q+s=m$ are fulfilled, it is easy to see that these integrals will give a non-zero result only when $l=0,1,2$ \cite{kn:sak}. In addition, for the integrals containing the products $Y_{1\pm 1}(\hat{\vec{v}})Y_{1\pm 1}(\hat{\vec{v}})$  the parameter $m$ has to be $m=\pm 2$ whereas it must be $m=0$ for the integrals with $Y_{1\pm 1}(\hat{\vec{v}})Y_{1\mp 1}(\hat{\vec{v}})$ to yield a non-zero contribution. However, although these integrals gives in principle a non-vanishing contribution, note that when we perform the integral over the variable $\hat{\vec{u}}$ in eq. (\ref{b.16}) the terms introduced by these contributions finally yield, by the orthogonality property of the spherical harmonics, a vanishing result. Thus, only when $l=0$ and $m=0$ contributes to $I_{3}$. Taking this into account and using eq. (\ref{b.15}) 

\begin{eqnarray}\label{b.18}
\nonumber I_{4}&=&\left(\frac{2\pi}{3}\right)^{3}\int_{0}^{t}dr\,\lambda(r)\left\{e^{-(2D_{r}-i\omega_{0})(t-r)}+e^{-(2D_{r}+i\omega_{0})(t-r)}\right\}\\
&&\times\int_{t}^{t+\tau}ds\,\lambda(s)\left\{e^{-(2D_{r}-i\omega_{0})(t+\tau-s)}+e^{-(2D_{r}+i\omega_{0})(t+\tau-s)}\right\}.
\end{eqnarray}

\noindent finally performing the changes of variables $t^{\prime}=t-r$ and $t^{\prime\prime}=t+\tau-s$ and using eq. (\ref{2.63}) we obtain

\begin{equation}\label{b.19}
I_{4}=\left(\frac{2\pi}{3}\right)^{3}\lambda^{2}(t)e^{-i\omega\tau}\left(\int_{0}^{\infty}dt\,e^{-i\omega t}\chi_{x}(t)\right)^{2}.
\end{equation}

\noindent In this integral the upper limit goes to infinity by causality reasons.

\newpage

\begin{figure}[htb]
\centerline{\psfig{file=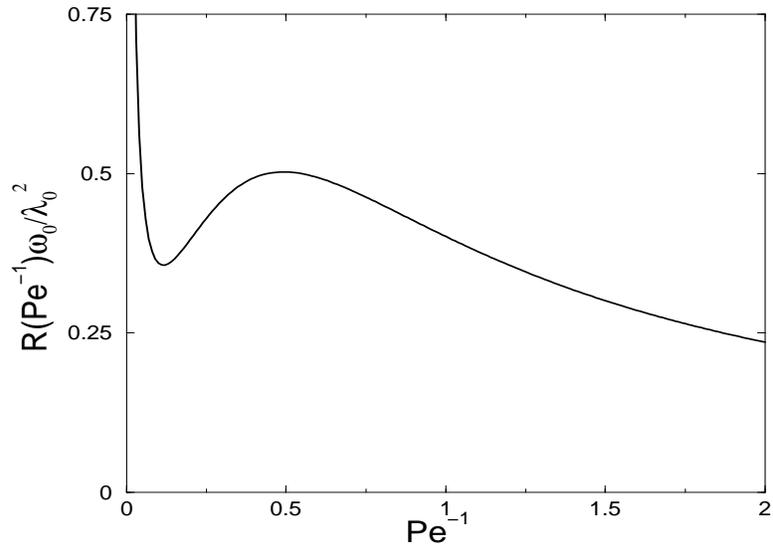,height=7.5cm,width=10cm}}
\caption{Signal to noise ratio as a function of the inverse of the P\'eclet number $Pe^{-1}=D_{r}/\omega_{0}$. We have represented non-dimensional quantities.}
%\label{fig:24}
\end{figure}

\begin{figure}[htb]
\centerline{\psfig{file=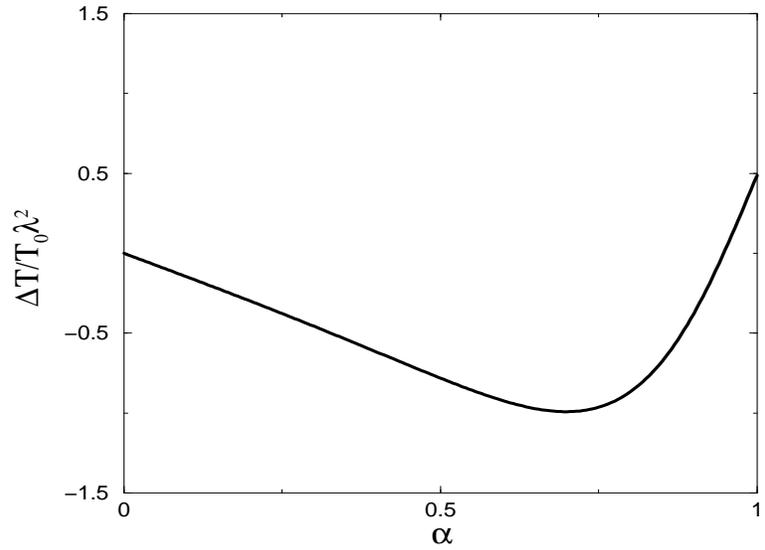,height=7.5cm,width=10cm}}
\caption{Mean first passage time as a function of the parameter $\alpha=\omega/\omega_{0}$. The presence of a minimum reveals the existence of stochastic resonance. $Pe^{-1}=D_{r}/\omega_{0}=0.08$. We have represented non-dimensional quantities.}
%\label{fig:24}
\end{figure}

\begin{figure}[htb]
\centerline{\psfig{file=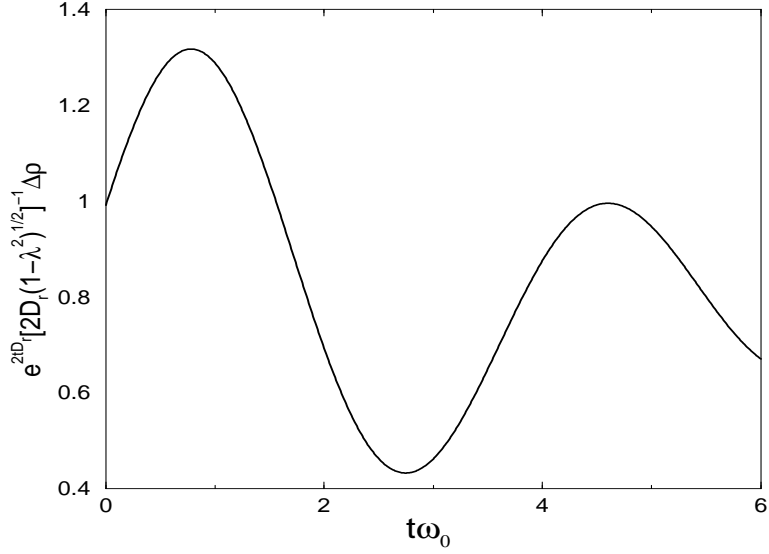,height=7.5cm,width=10cm}}
\caption{Escape time distribution for a dipole in a shear flow under an oscillating magnetic field. The succession of maxima is a signature of the presence of stochastic resonance in this system. $Pe^{-1}=D_{r}/\omega_{0}=0.08$ and $\alpha=\omega/\omega_{0}=0.7$. We have represented non-dimensional quantities.}
%\label{fig:24}
\end{figure}

\begin{figure}[htb]
\centerline{\psfig{file=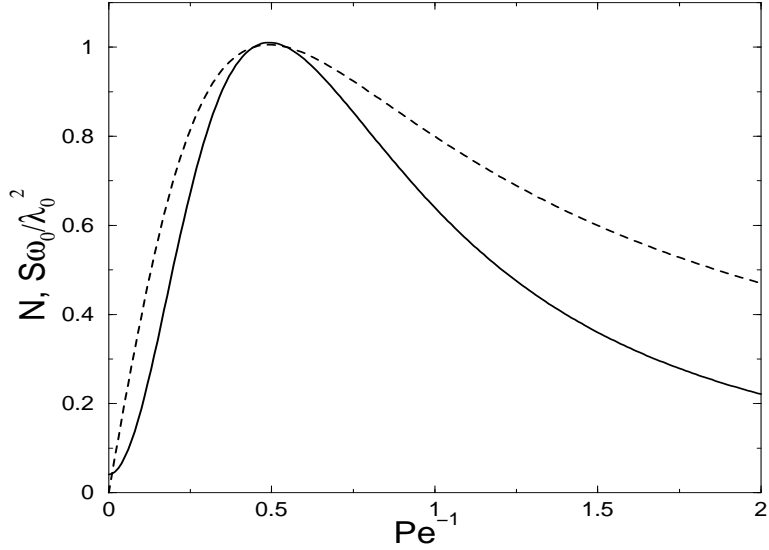,height=7.5cm,width=10cm}}
\caption{Output signal and output background signal as function of the inverse of the P\'eclet number. Solid line represents the quantity $S(Pe^{-1})\omega_{0}/\lambda_{0}^{2}$ whereas dashed line represents $N(Pe^{-1})$. We have taken $\alpha=0.1$. We have represented non-dimensional quantities.}
%\label{fig:24}
\end{figure}

\begin{figure}[htb]
\centerline{\psfig{file=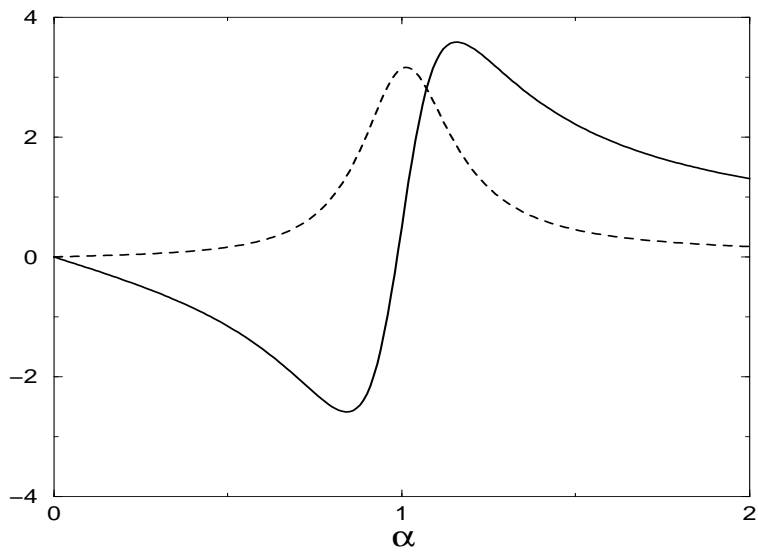,height=7.5cm,width=10cm}}
\caption{Comparison between the imaginary part of the susceptibility of the signal $\hat{R}_{x}$ (solid line) and the noisy part of the spectrum computed from the Fokker-Planck equation (dashed line). $Pe^{-1}=D_{r}/\omega_{0}=0.08$. We have represented non-dimensional quantities.}
%\label{fig:24}
\end{figure}

\end{document}